\documentclass[showpacs,pra,a4paper,floatfix,superscriptaddress,twocolumn]{revtex4}
\usepackage{graphicx}
\usepackage{bm,epsfig}
\usepackage[dvips]{pstcol}
\usepackage{amsmath}
\usepackage{amssymb}

\begin{document}

\title{Light propagation through closed-loop atomic media\\
beyond the multiphoton resonance condition}

\author{Mohammad \surname{Mahmoudi}}
\email{mahmoudi@iasbs.ac.ir}
\affiliation{Max-Planck-Institut f\"ur Kernphysik, Saupfercheckweg 1, 
D-69117 Heidelberg, Germany}
\affiliation{Physics Department, Zanjan University, P. O. Box 45195-313,
Zanjan, Iran}

\author{J\"org \surname{Evers}}
\email{joerg.evers@mpi-hd.mpg.de}
\affiliation{Max-Planck-Institut f\"ur Kernphysik, Saupfercheckweg 1, 
D-69117 Heidelberg, Germany}

\date{\today}

\begin{abstract}
The light propagation of a probe field pulse in a four-level 
double-lambda type system driven by laser fields that form
a closed interaction loop is studied. Due to the finite frequency
width of the probe pulse, a time-independent analysis relying
on the multiphoton resonance assumption is insufficient.
Thus we apply a Floquet decomposition of the equations of
motion to solve the time-dependent problem beyond the
multiphoton resonance condition.
We find that the various Floquet components can be interpreted
in terms of different scattering processes, and that the 
medium response oscillating in phase with the probe field in 
general is not phase-dependent. The phase dependence
arises from a scattering of the coupling fields into the
probe field mode at a frequency which in general differs 
from the probe field frequency.
We thus conclude that in particular for short pulses with 
a large frequency width, inducing a closed loop 
interaction contour may not 
be advantageous, since otherwise the phase-dependent medium
response may lead to a distortion of the pulse shape.
Finally, using our time-dependent analysis, we demonstrate 
that both the closed-loop and the non-closed loop configuration 
allow for sub- and superluminal light propagation with small
absorption or even gain.
Further, we identify one of the coupling field Rabi frequencies 
as a control parameter that allows to conveniently switch between
sub- and superluminal light propagation.
\end{abstract}


%
%
%

\maketitle

\section{\label{intro}Introduction}

The propagation of a light pulse through a dispersive medium has
been extensively investigated~\cite{Harris,Schmit,Field,Xiao}.
Recently the amplitude and the phase control of the group velocity
in a transparent media have attracted much attention. It is well
known that the propagation velocity of a light pulse can be slowed 
down~\cite{Hau,Kash}, can become greater than  the light propagation
velocity in vacuum, or even negative in transparent 
media~\cite{Steinberg,Wang}. Superluminal light propagation is a
phenomenon in which the group velocity of an optical pulse in a
dispersive medium is greater than of the light in vacuum. Despite the
name superluminal light, it is generally believed that no information
can be sent faster than light speed $c$ in vacuum as explained by
Chiao~\cite{Chiao}. Thus, a group velocity
faster than $c$ as reported here does not violate Einstein principle 
of special
relativity. The superluminal light propagation has been
investigated for many potential uses, not only as a tool for
studying a very peculiar state of matter, but also for developing
quantum computers, high speed optical switches and communication
systems~\cite{Kim}.

There have been only few experimental and theoretical studies
which realized both superluminal and subluminal light propagation in a
single system. Talukder et al. have shown femtosecond laser pulse
propagation from superluminal to subluminal velocities in an absorbing
dye by changing the dye concentration~\cite{Talukder}. Shimizu et al.
were also able to control a light pulse speed, with only a few
cold atoms in a high-finesse microcavity by detuning the laser
frequency  from a cavity resonance frequency-locked to the atomic
transition~\cite{Shimizu}. 
Most schemes for a speed control in atomic
systems involve changing the frequency, amplitude or
phase difference of the applied fields. It was shown that switching
from subluminal to superluminal pulse propagation can be achieved
via the intensity of coupling
field~\cite{Goren,Agarwal,Han,Tajalli}, and via the relative phase between
two weak probe fields~\cite{Bortman}. The intensity control of
group velocity has been attempted by using a lower level-coupling
field in the three-level atomic system~\cite{Agarwal}. In  
another study, a scheme based on  four-level electromagnetically
induced transparency for switching from subluminal to superluminal
light propagation via relative phase between driving fields was
introduced~\cite{Sahrai}. Recently we have used an incoherent pump
field to control the light propagation from subluminal to
superluminal~\cite{Mahmoudi1, Mahmoudi2}. Further, we have
studied the propagation of light pulses in three-level closed
laser-interaction loop atomic systems, where we exploited the fact 
that changing the relative phase between applied fields can modify the
absorption and dispersion properties of the medium~\cite{Mahmoudi3}.
The double-$\Lambda$ setup is another scheme which allows for a closed
laser-field interaction loop and provides a very rich spectrum of 
phenomena based
on atomic coherence. One reason for this are the various interfering 
excitation channels in such a system~\cite{Korsunsky}. The
double $\Lambda$ system has been investigated in the content of
amplification without inversion~\cite{Kocharovskaya}, phase
sensitive laser cooling~\cite{Kosachiov}, the propagation of the
pairs of optical pulses~\cite{Gerboneschi}, optical phase
conjugation~\cite{Lukin}, phase control of 
photoionization~\cite{Li}, phase control of electromagnetically induced
transparency~\cite{Korsunsky} and coherent population trapping
\cite{Maichen}. Recently Morigi et al~\cite{Morigi} have compared
the phase-dependent properties of the $\diamond$ (diamond) four
level system with those of the double $\Lambda$ system.
The propagation of light in a double $\Lambda$ medium was studied
earlier in Ref~\cite{Kocharovskaya}. The authors considered the
case of complete resonance and the coherent population trapping
(CPT) condition initially
fulfilled, and found the possibility of amplification of one of
the frequency pairs. Korsunsky et al.
present a theory of a continuous wave light propagation in a
medium of atoms with a double $\Lambda$ configuration~\cite{Korsunsky2}. 
They have shown that, when the so-called  multiphoton
resonance condition is fulfilled, both absorptive and dispersive
properties of a such a medium depends on the relative phase of the
driving fields. 

\begin{figure}[t]
\includegraphics[width=8cm]{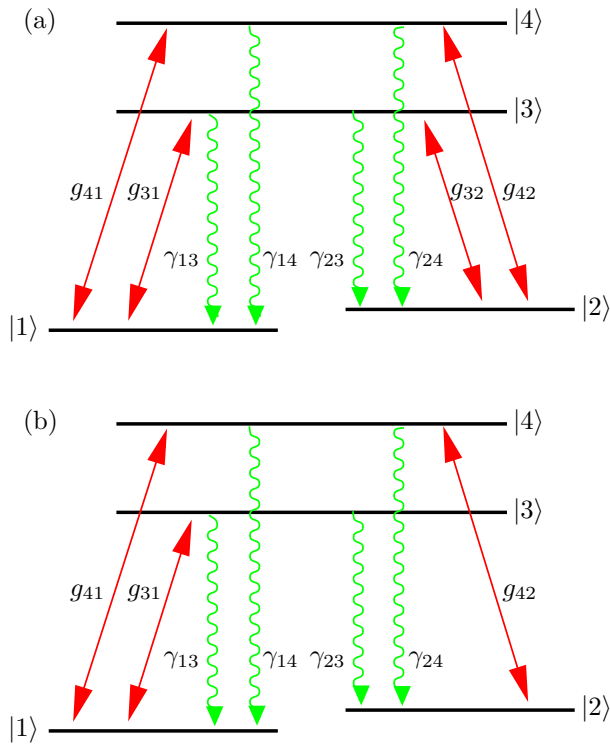}
\caption{\label{fig-system}(Color online)
The four-level double-$\Lambda$ type schemes considered 
in the analysis.
(a) is driven by four laser fields indicated by the
red arrows with Rabi frequencies
$g_{ij}$ ($i\in \{3,4\}$,$j\in \{1,2\}$) that form a closed
interaction loop. This loop gives rise to a dependence
on the relative phase of the different laser fields, and
in general makes a time-independent steady state of the system
dynamics impossible.
(b) is the corresponding level scheme where one of the driving 
fields has been removed ($g_{32}=0$). Here, the laser fields
do not form a closed loop.
The spontaneous decays with rates $\gamma_{ij}$ 
($i\in \{1,2\}$,$j\in \{3,4\}$) are denoted by the wiggly green lines.
}
\end{figure}

A peculiarity of the closed laser interaction loop systems is the
fact that a certain initial atomic state is connected to another
atomic state via several combinations of laser field interactions.
On the one hand, this gives rise to a dependence of the optical 
properties of the system on the relative
phase of the driving fields. This allows for a great control
of the optical properties and has been exploited in various
circumstances, as discussed above. The drawback, however, is
that such a system in general does not have a time-independent
steady state. A time-independent steady state is reached only 
if a particular linear combination of
the detunings of all incident laser fields is zero, that is, if
the so-called multiphoton resonance condition is fulfilled.
This assumption was made in the previous studies.
But for a meaningful definition of the group velocity of a probe 
pulse in a given medium, one has to require that the medium 
itself including the coupling fields, which in this context 
merely aid in the preparation of the medium, has properties 
independent of the probe pulse characteristics. 
While the probe pulse which is finite in time necessarily
consists of different frequency components and thus interacts with the
atom with different detunings, the coupling fields have to be kept fixed 
in the calculation of the probe field propagation, since the different
frequency components interact with the medium simultaneously.
In particular, it is not possible to adjust one of the coupling field
detunings such that the multiphoton resonance condition is maintained
throughout the pulse propagation.
Thus, it is impossible to assume the multiphoton resonance
condition in the evaluation of the propagation of a finite
probe pulse, and a time-dependent analysis is required.

Therefore in this paper, we investigate  the light 
propagation in a four-level double lambda system
as shown in Fig.~\ref{fig-system}(a)
both with and without assuming the multiphoton resonance condition. 
We apply laser fields to all four dipole-allowed transitions
to generate a closed laser-interaction loop.
In the time-independent case where the multiphoton 
resonance condition is fulfilled, we find sub- and 
superluminal light propagation, controlled simply by the 
relative phase of the driving fields.
For the time-dependent study beyond the multiphoton resonance
condition, we apply a Floquet decomposition to the in general 
time dependent equations of motion. By comparing the Floquet 
decomposition to the time-independent treatment, we identify 
the respective frequency components found in this expansion 
with different interaction pathways in the closed loop system. 
We find that the medium response oscillating in phase with the 
probe field in general is not phase-dependent. The phase-dependent
process contributing to the probe field susceptibility
only occurs at a specific frequency, and is thus likely
to distort the shape of a short probe pulse considerably if
this specific frequency is contained in the wide 
frequency spectrum of the short pulse.
Thus we conclude that for parameters violating the multiphoton
resonance condition as in the probe pulse propagation, 
inducing a closed loop interaction contour may
not be advantageous. Nevertheless, we demonstrate that also under
conditions without a closed loop as shown in Fig.~\ref{fig-system}(b), 
both sub- and superluminal light propagation is possible with gain.
Finally, we identify the Rabi frequency of one of the coupling fields
as a convenient control parameter which allows to switch
between sub- and superluminal light propagation.

\section{\label{theory}Theoretical Analysis}
\subsection{The Model}

We consider a four-level atomic systems in a double-$\Lambda$
configuration as depicted in Figure~\ref{fig-system}. The scheme consists
of two metastable lower states $|1\rangle$, $|2\rangle$ and 
two excited states $|3\rangle$, $|4\rangle$. The electric-dipole 
allowed transitions $|1\rangle-|3\rangle$, $|2\rangle-|3\rangle$ and
$|2\rangle-|4\rangle$ are driven by three coherent laser fields,
while a weak tunable coherent probe field with central frequency
$\omega_p=\omega_{41}$ 
is applied to the dipole-allowed transition $|1\rangle-|4\rangle$. 
The spontaneous decay rates from level $|i\rangle$ ($i\in \{3,4\}$)
to the levels $|j\rangle$ ($j\in \{1,2\}$) are denoted
by $2 \gamma_{ji}$. 
The electromagnetic driving fields applied to transition
$|i\rangle-|j\rangle$ can be written as
\begin{align}
\label{field}
\bm{E}_{ij} = E_{ij}\hat{\bm{e}}_{ij} 
e^{-i(\omega_{ij}t-\vec{k}_{ij} \vec{r} +\phi_{ij})} + \textrm{ c.c.}\,,
\end{align}
with amplitude $E_{ij}$, unit polarization vectors
$\hat{\bm{e}}_{ij}$, frequencies $\omega_{ij}$, wave vectors $\vec{k}_{ij}$
and absolute phase $\phi_{ij}$. 
Note that we do not assume absolute phase control in our calculations,
but only relative phase control between the different driving fields.

The Hamiltonian in dipole and rotating wave approximation is given 
by~\cite{scullybook,FiSw2005}
\begin{align}
H = &\sum_{j=1}^{4} E_j|j\rangle\langle j| \nonumber \\
&- \sum_{l=3}^{4}\sum_{m=1}^{2} \left (
\hbar g_{lm} e^{-i\alpha_{lm}} 
 |l\rangle\langle m | + \textrm{ h.c.} \right)\,,
\end{align}
where the corresponding Rabi frequencies are denoted 
by $g_{ij}=E_{ij} (\hat{\bm{e}}_{ij} \vec{d}_{ij})/\hbar$ with
$\vec{d}_{ij}$ as the
atomic dipole moment of the  corresponding transition. 
The energies of the involved states are denoted 
$E_i$ ($i\in\{1,\dots,4\}$), and we define the 
transition frequencies $\bar{\omega}_{ij}=(E_i-E_j)/\hbar$.
The arguments of the exponential functions are given by
\begin{align}
\label{alpha}
\alpha_{ij} = \omega_{ij}t-\vec{k}_{ij} \vec{r} +\phi_{ij} \,.
\end{align}
In a suitable reference frame, the Hamiltonian can be
written as
\begin{align}
\label{ham-int}
V = & \hbar (\Delta_{32}-\Delta_{31})\tilde{\rho}_{22}
- \hbar \Delta_{31} \tilde{\rho}_{33}
\nonumber \\
&+\hbar (\Delta_{32}-\Delta_{31}-\Delta_{42})\tilde{\rho}_{44}
 \nonumber \\
&- \hbar \left (g_{31} \tilde{\rho}_{31}
+ g_{32} \tilde{\rho}_{32} 
+ g_{42} \tilde{\rho}_{42} \right. \nonumber \\
& \left. + g_{41} \tilde{\rho}_{41} e^{-i\Phi} 
 + \textrm{ h.c.}\right)\,.
\end{align}
Here, we have defined $\rho_{ij} = |i\rangle \langle j|$ and
the corresponding operator in the new reference frame
is denoted by $\tilde{\rho}_{ij}$ ($i,j\in\{1,\dots,4\}$).
It is interesting to note that now the residual time dependence
along with the laser field phases appears only together
with the probe field Rabi frequency $g_{41}$ in the
parameter $\Phi$ given by
\begin{subequations}
\label{phi}
\begin{align}
\Phi &= \Delta t-\vec{K}\vec{r}+\phi_0 \,, \\
\Delta &= (\Delta_{32}+\Delta_{41})-(\Delta_{31}+\Delta_{42}) \,,\\
\vec{K} &= (\vec{k}_{32}+\vec{k}_{41})-(\vec{k}_{31}+\vec{k}_{42}) \,,\\
\phi_0 &= (\phi_{32}+\phi_{41})-(\phi_{31}+\phi_{42})\,.
\end{align}
\end{subequations}
The parameters $\Delta$, $\vec{K}$ and $\phi_0$
are known as the multiphoton resonance detuning, wave vector 
mismatch and initial phase difference, respectively.
Note that in general it is not possible to find a reference frame
where the explicit time dependence due to $\Delta$ vanishes
from the Hamiltonian, such that for $\Delta\neq 0$ no stationary
long-time limit can be expected.
Using the notations Eqs.~(\ref{phi}) and Eq.~(\ref{alpha}), 
the transformation
of the operator $\rho_{41}$, which will be of particular 
interest later on, to the chosen interaction 
picture  can be written as 
\begin{align}
\label{trafo-41-a}
\tilde{\rho}_{41} &=  e^{-i\Phi}\,e^{i\alpha_{41}}\,\rho_{41}\,.
\end{align}
We further define $\hat{\rho}_{41}$ as the coherence $\rho_{41}$
in a reference frame oscillating in phase with the probe field
$\bm{E}_{41}$ as
\begin{align}
\label{trafo-41-b}
\hat{\rho}_{41} &=  e^{i\alpha_{41}}\,\rho_{41} = 
  e^{i\Phi}\tilde{\rho}_{41}\,.
\end{align}

From the Hamiltonian Eq.~(\ref{ham-int}), and including the spontaneous decay
in Born-Markov approximation, the density matrix equations
of motion can be derived as
\begin{subequations}
\label{density}
\begin{align}
\frac{\partial}{\partial t} \tilde{\rho}_{11} &= i g^{\ast}_{31} \tilde{\rho}_{31} - i g_{31}
\tilde{\rho}_{13} + i g^{\ast}_{41} \tilde{\rho}_{41} e^{i\Phi}  \nonumber  \\
   & -i g_{41} \tilde{\rho}_{14}e^{-i\Phi}+2
     \gamma_{13}\tilde{\rho}_{33}+2\gamma_{14}\tilde{\rho}_{44} \,,  \\
\frac{\partial}{\partial t} \tilde{\rho}_{22}&= i g^{\ast}_{32} \tilde{\rho}_{32}-i g_{32}
   \tilde{\rho}_{23}+i g^{\ast}_{42} \tilde{\rho}_{42}  \nonumber \\
   &-i g_{42} \tilde{\rho}_{24}+2 \gamma_{23}\tilde{\rho}_{33}+2\gamma_{24}\tilde{\rho}_{44} \,,
    \allowdisplaybreaks[2] \\
\frac{\partial}{\partial t} \tilde{\rho}_{33}&=-i g^{\ast}_{31} \tilde{\rho}_{31}-i g^{\ast}_{32}
  \tilde{\rho}_{32}+i g_{31} \tilde{\rho}_{13}\nonumber \\
  & +i g_{32} \tilde{\rho}_{23}-2 \gamma_{3}\tilde{\rho}_{33} \,, 
   \allowdisplaybreaks[2]\\
\frac{\partial}{\partial t} \tilde{\rho}_{12}&= i(\Delta_{32}-\Delta_{31})\tilde{\rho}_{12}+i g^{\ast}_{31}
  \tilde{\rho}_{32}-i g_{32} \tilde{\rho}_{13} \nonumber \\
  &+i g^{\ast}_{41} \tilde{\rho}_{42}e^{i\Phi} -i g_{42}
  \tilde{\rho}_{14}-\Gamma_{12}\tilde{\rho}_{12} \,, 
  \allowdisplaybreaks[2]\\
\frac{\partial}{\partial t} \tilde{\rho}_{13}&= -i\Delta_{31}\tilde{\rho}_{13}+i g^{\ast}_{31}(
  \tilde{\rho}_{33}-\tilde{\rho}_{11})-i g^{\ast}_{32} \tilde{\rho}_{12} \nonumber \\
  &+i g^{\ast}_{41} \tilde{\rho}_{43}e^{i\Phi} -\Gamma_{13}\tilde{\rho}_{13} \,, 
  \allowdisplaybreaks[2]\\
\frac{\partial}{\partial t} \tilde{\rho}_{14}&=
  i(\Delta_{32}-\Delta_{31}-\Delta_{42})\tilde{\rho}_{14}+i g^{\ast}_{41}e^{i\Phi}
  (\tilde{\rho}_{44}-\tilde{\rho}_{11}) \nonumber \\
  & -i g^{\ast}_{42} \tilde{\rho}_{12}+i g^{\ast}_{31} \tilde{\rho}_{34}-\Gamma_{14}\tilde{\rho}_{14} \,, 
  \allowdisplaybreaks[2]\\
\frac{\partial}{\partial t} \tilde{\rho}_{23}&= -i\Delta_{32}\tilde{\rho}_{23}+i g^{\ast}_{32}(
  \tilde{\rho}_{33}-\tilde{\rho}_{22})-i g^{\ast}_{31} \tilde{\rho}_{21} \nonumber \\
 & +i g^{\ast}_{42} \tilde{\rho}_{43}-\Gamma_{23}\tilde{\rho}_{23} \,, 
 \allowdisplaybreaks[2]\\
\frac{\partial}{\partial t} \tilde{\rho}_{24}&= -i\Delta_{42}\tilde{\rho}_{24}+i g^{\ast}_{42}(
  \tilde{\rho}_{44}-\tilde{\rho}_{22})-i g^{\ast}_{41} \tilde{\rho}_{21}e^{i\Phi} \nonumber \\
  &+i g^{\ast}_{32} \tilde{\rho}_{34}-\Gamma_{24}\tilde{\rho}_{24} \,, 
  \allowdisplaybreaks[2] \\
\frac{\partial}{\partial t} \tilde{\rho}_{34}&= -i(\Delta_{42}-\Delta_{32})\tilde{\rho}_{34}+i
  g_{31} \tilde{\rho}_{14}+i g_{32} \tilde{\rho}_{24} \nonumber \\
  & -i g^{\ast}_{41} \tilde{\rho}_{31}e^{i\Phi} -i g^{\ast}_{42}
  \tilde{\rho}_{32}-\Gamma_{34}\tilde{\rho}_{34} \,, 
  \allowdisplaybreaks[2]\\
\tilde{\rho}_{44}&= 1-\tilde{\rho}_{11}-\tilde{\rho}_{22}-\tilde{\rho}_{33} \,. 
\end{align}
\end{subequations}
We have further defined $\gamma_j=\gamma_{1j}+\gamma_{2j}$,
and our chosen level scheme implies $\gamma_1=\gamma_2=0$.
$\Gamma_{ij}=(2\gamma_i+2\gamma_j)/2$ ($i\in\{1,2\}$ and $j\in\{3,4\}$) 
are the damping rates of the coherences on
transitions $|i\rangle-|j\rangle$.

Due to the explicit time dependence of Eqs.~(\ref{density})
via the parameter $\Delta$ in $\Phi$, it 
is clear that in general
the system does not have a constant steady state solution.
If, however, the so-called multiphoton resonance condition
$\Delta=0$, $\vec{K}=0$ is fulfilled, then the equations of 
motion (\ref{density}) have constant coefficients and thus a 
stationary solution in the
long-time limit can be found, which for a suitable choice
of parameters depends on the constant 
relative phase $\phi_0$.

\subsection{\label{suscep}Linear susceptibility and group velocity}

The linear susceptibility of the weak probe field can be written
as~\cite{scullybook,FiSw2005}
\begin{equation}
\label{sus}
\chi (\omega_p) = \frac{2N d_{41}}{\epsilon_0
E_{41}}\rho_{41}(\omega_p) \,,
\end{equation}
where $N$ is the atom number density in the medium and
$\chi=\chi^\prime+i\chi^{\prime\prime}$. The real and imaginary
parts of $\chi(\omega_p)$ correspond to the dispersion and the
absorption respectively. This expression for the susceptibility
usually is applied to a single probe field frequency $\omega_p$ 
of a narrow-bandwidth continuous-wave laser field. 
If the probe field consists of short pulses such that the individual 
probe pulses have a non-negligible frequency width, then Eq.~(\ref{sus})
describes the propagation of an individual frequency component
of the probe pulse.

For a realistic example, we consider
the sodium $D_1$ transition, in which the transition rate, dipole
moment and atom density are $\gamma=2\pi\times 9.76$ MHz,
$d_{41}=2.1\times 10^{-29}$ Cm and $N = 1.3\times 10^{12}$
cm$^{-3}$, respectively. The probe field with $g_{41}=0.01\gamma$
is applied to the system, leading to
$(2\,N\,d_{41})/(\epsilon_0 E_{41})\simeq 1$.
Eq.~(\ref{sus}) refers to the part of the coherence
$\rho_{41}(\omega_p)$ oscillating at a frequency of the incident probe 
beam. This coincides with the transformed coherence $\hat{\rho}_{41}$
introduced in Eq.~(\ref{trafo-41-b}). 
Thus for our choice of example parameters, we obtain
\begin{equation}
\label{observable}
\chi (\omega_p) \simeq \hat{\rho}_{41}\,.
\end{equation}
Therefore throughout our numerical study, we will
discuss the transformed coherence $\hat{\rho}_{41}$
in order to be study the light propagation in the medium.

We also introduce the so-called 
group index, $n_g=c/v_g$ where $v_g$, the group velocity of the
probe field, is given by~\cite{scullybook,FiSw2005}
\begin{equation}
v_g=\frac{c}{1+2\pi\chi^\prime
(\omega_p)+2\pi\frac{\partial}{\partial
\omega_p}\chi^\prime(\omega_p)}
\end{equation}
This expression shows that, for a small imaginary part
$\chi^{\prime\prime}(\omega_p)$ and positive steep dispersion, the group
velocity is significantly reduced. On the other hand, strong
negative dispersion can leads to an increase in the group velocity
and even to a negative group velocity.


\subsection{Multi-photon resonance approximation}

In this section, we assume the multi-photon resonance condition
$\Delta=0$, $\vec{K}=0$ to be fulfilled. Then, the coefficients of
the density matrix equations Eqs.~(\ref{density}) do not have an explicit 
time dependence, and the system has a stationary steady state.
In the particular case
\begin{subequations}
\label{simplepar}
\begin{align}
\Delta_{31}&=\Delta_{32}=\Delta_{41}=\Delta_{42}=0\,, \\
\gamma &= \gamma_{14} = \gamma_{24} =\gamma_{13} =\gamma_{23} \,,
\end{align}
\end{subequations}
a simple analytical expression for the steady state solution 
of $\tilde{\rho}_{41}$ up to leading order in $g_{41}$
can be found:
\begin{figure*}[t]
\includegraphics[width=14cm]{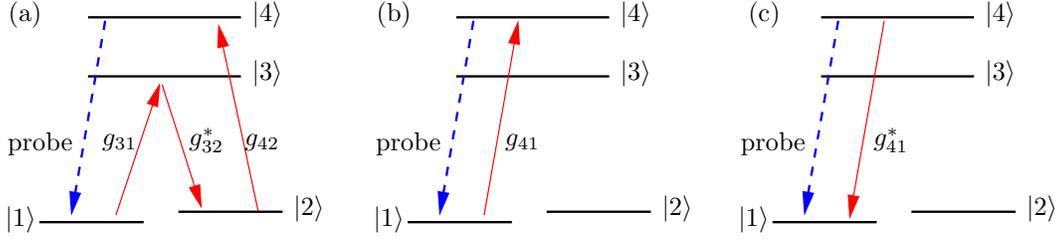}
\caption{\label{fig-path}(Color online) Interpretation of the
different contributions to the probe field susceptibility
in terms of transition pathways. (a) represents the
interaction loop leading to a scattering of the driving
fields into the probe field mode. (b) is the direct
scattering of the probe field off of the probe transition.
(c) shows a counter-rotating term. The solid red arrows
indicate coupling field transitions, the dashed blue line
is a probe field interaction.}
\end{figure*}
\begin{align}
\tilde{\rho}_{41} = & \frac{- i\, \gamma\, g_{32}^\ast \, g_{31}\, g_{42}}
{2 \, D} \nonumber \\[2ex] 
&+ \frac{i \gamma^3 \, (|g_{31}|^2+|g_{32}|^2+|g_{42}|^2)
\, |g_{32}|^2 \, g_{41} e^{-i \phi_0}}
{2 \, D^2} \nonumber \\[2ex] 
&- \frac{i \gamma \, |g_{31}|^2 \, |g_{32}|^2 \, |g_{42}|^2
\, g_{41}^\ast e^{i \phi_0}}
{2 \, D^2}  \,, 
\end{align}
where
\begin{align}
D = & 
|g_{31}|^2\cdot |g_{42}|^2  
+\gamma^2 \left( |g_{31}|^2 
 + |g_{32}|^2 + |g_{42}|^2 \right)\,. \label{def-d}
\end{align}
Transferring to the frame rotating in phase
with the probe field using
Eqs.~(\ref{trafo-41-a}) and (\ref{trafo-41-b}), we obtain
\begin{align}
\label{ana}
\hat{\rho}_{41} =& \frac{- i\, \gamma\, g_{32}^\ast \, g_{31}\, g_{42}\,e^{i\phi_0}}
{2 \, D} \nonumber \\[2ex] 
&+ \frac{i \gamma^3 \, (|g_{31}|^2+|g_{32}|^2+|g_{42}|^2)
\, |g_{32}|^2 \, g_{41} }
{2 \, D^2} \nonumber \\[2ex] 
&- \frac{i \gamma \, |g_{31}|^2 \, |g_{32}|^2 \, |g_{42}|^2
\, g_{41}^\ast e^{2 i \phi_0}}
{2 \, D^2}  \,.
\end{align}
First, we note that the whole expression for $\hat{\rho}_{41}$ 
oscillates with same frequency as the probe field, as evidenced
by the lack of an overall time-dependent factor in $\hat{\rho}_{41}$, 
and therefore
contributes to the probe field susceptibility.
The three contributions to Eq.~(\ref{ana}) admit a simple interpretation
in terms of the  involved physical processes.
The first component is proportional to 
$g_{32}^\ast \, g_{31}\, g_{42}$, and thus corresponds to a 
closed interaction loop involving a sequence of all four dipole-allowed
transitions, or  a scattering of
the driving field modes into the probe field mode.
This round-trip depends on the relative multiphoton
phase $\phi_0$, which therefore occurs in this contribution.
The second term in Eq.~(\ref{ana}) 
proportional to $g_{14}$ represents a direct scattering
of the probe field off of the probe transition, 
involving an excitation and a de-excitation of the probe transition.
Since no closed interaction-contour via different laser fields
is involved, this contribution
does not depend on the relative phase $\phi_0$.
The last contribution in Eq.~(\ref{ana})
proportional to $g_{14}^\ast$ depends on twice the relative 
field phase $\phi_0$ can be interpreted as a counter-rotating 
contribution.
This preliminary interpretation of the individual contributions 
will become more transparent in the time-dependent analysis without
the assumption of multiphoton resonance in the following
Sec.~\ref{beyond}.


\subsection{\label{beyond}Beyond the multi-photon resonance approximation}

We now drop the multi-photon resonance condition and evaluate
the equations of motion Eqs.~(\ref{density}) for the general
case of time-dependent coefficients. The Eqs.~(\ref{density}) 
can be written in a compact form
\begin{equation}\label{R}
  \frac{\partial}{\partial t} \tilde{R}+\Sigma=M \tilde{R},
\end{equation}
where
\begin{align}
\tilde{R}  = & ( \tilde{\rho}_{11},\tilde{\rho}_{12},\tilde{\rho}_{13},
\tilde{\rho}_{14},\tilde{\rho}_{21},\tilde{\rho}_{22},
\tilde{\rho}_{23},\tilde{\rho}_{24}, \nonumber \\
& 
\tilde{\rho}_{31},\tilde{\rho}_{32},\tilde{\rho}_{33},
\tilde{\rho}_{34},\tilde{\rho}_{41},\tilde{\rho}_{42},
\tilde{\rho}_{43})^T 
\end{align}
is a vector containing the density matrix elements, and
\begin{align}
\Sigma = & ( -2\gamma_{14}, 0, 0, -i\bar{g}^\ast_{41}e^{i\Delta t}, 0, 
-2\gamma_{24}, \nonumber \\
& 0, -i g^\ast_{42},  0 ,  0 ,  0, 0, i\bar{g}_{41}e^{-i\Delta t}, ig_{42},
0)^T 
\end{align}
is a vector independent of the density matrix elements which arises
from eliminating one of the state populations from the equations
of motion via the trace condition $Tr(\tilde{\rho}) = 1$.
Note that we have introduced the notation 
\begin{align}
\bar{g}_{41} = g_{41}\, e^{-i \vec{K}\vec{r} + i\phi_0}\,.
\end{align}
The matrix $M$ follows from Eqs.~(\ref{density}).
Both the matrix $M$ and the $\Sigma$ can be separated into terms
with different time dependence~\cite{Hu},
\begin{subequations}
\begin{align}
\label{decomposition}
\Sigma &=\Sigma_0+\bar {g}_{41}\Sigma_1 e^{-i\Delta t}
+\bar{g}^{\ast}_{41}\Sigma_{-1}e^{i\Delta t} \,, \\
M &= M_0+\bar{g}_{41}M_1 e^{-i\Delta t}+\bar{g}^{\ast}_{41}
M_{-1} e^{i\Delta t}\,,
\end{align}
\end{subequations}
where $\Sigma_0, \Sigma_{\pm 1}, M_0$ and $M_{\pm 1}$ are time
independent.
Using these definitions in Eq.~(\ref{R}), we obtain
\begin{align}
\label{R1}
& \frac{\partial}{\partial t} \tilde{R} +\Sigma_0+
\bar {g}_{41} e^{-i\Delta t}\Sigma_{+1}
+\bar {g}^\ast_{41}e^{i\Delta t} \Sigma_{-1}
\nonumber \\
& =(M_0+\bar {g}_{41}e^{-i\Delta t}M_1+\bar {g}^\ast_{41}
e^{i\Delta t}M_{-1} ) \tilde{R} \,.
\end{align}
It is important to note that the time dependence of Eqs.~(\ref{density})
only arises from the parameter $\Delta$, such that the coefficients
are periodic in time. According to Floquet's theorem~\cite{floquet}, 
the solution 
$\tilde{R}$ therefore has only contributions oscillating at harmonics of 
the detuning $\Delta$. The higher-order harmonics in this
frequency expansion are suppressed by powers of the probe field 
Rabi frequency $g_{41}$ relative to the
other frequencies involved in the system.
Since we are interested in the
case where the probe field is weak, we truncate the Floquet expansion 
after the leading first order and obtain as ansatz for the 
solution $\tilde{R}$:
\begin{equation}
\label{R2}
\tilde{R}=\tilde{R}_0+\bar {g}_{41} e^{-i\Delta t}\tilde{R}_1
+\bar {g}_{41}^\ast e^{i\Delta t}\tilde{R}_{-1}\,.
\end{equation}
By using Eq.~(\ref{R2}) in Eq.~(\ref{R1}) and equating the
coefficients oscillating at different harmonics of $\Delta$, 
we obtain the solutions for $\tilde{R}_0$ and $\tilde{R}_{\pm 1}$ as
\begin{subequations}
\label{floq-solution}
\begin{eqnarray}
\tilde{R}_0 &=& M^{-1}_0 \Sigma_0 \,, \\
\tilde{R}_1 &=& (M_0+i\Delta)^{-1}(\Sigma_{1}-M_1 \tilde{R}_0) \,, \\
\tilde{R}_{-1} &=& (M_0-i\Delta)^{-1}(\Sigma_{-1}-M_{-1} \tilde{R}_0) \,.
\end{eqnarray}
\end{subequations}
The medium response is determined by the 13th component
$\tilde{\rho}_{41}$ of $\tilde{R}$. Transferring back
$\tilde{\rho}_{41}$ to the operator
$\hat{\rho}_{41}$ in the frame oscillating in phase with the
probe field using Eqs.~(\ref{trafo-41-a}) and (\ref{trafo-41-b}), we obtain
\begin{align}
\label{rho-41}
\hat{\rho}_{41} =  [\tilde{R}_0]_{13}\,e^{i\Phi} &+ g_{41} 
[\tilde{R}_1]_{13}
 + g_{41}^\ast [\tilde{R}_{-1}]_{13}\,e^{2 i\Phi} \,.
\end{align}
Here, $[x]_{13}$ denotes the 13th component of $x$.
It can be seen that the contribution proportional to 
$[\tilde{R}_1]_{13}$ oscillates in phase with the probe beam, and
thus contributes to the probe beam susceptibility independent
of the frequency of the incident driving field.
For $\Delta \neq 0$, the two other contributions proportional
to $[\tilde{R}_0]_{13}$ and $[\tilde{R}_{-1}]_{13}$ oscillate at different
frequencies due to the time dependence of $\Phi$ and thus 
do not contribute to the probe beam 
susceptibility. However, even in the case $\Delta \to 0$, 
the term $[\tilde{R}_1]_{13}$ is distinct in that the two other
contributions in general propagate differently due 
to the wave vector
mismatch $\vec{K}$.

Using the result Eq.~(\ref{rho-41}), we are now also in
the position to in detail understand the expression 
Eq.~(\ref{ana}) we
previously derived assuming $\Delta = 0$ and $\vec{K}=0$.
For this, we explicitly evaluate $[\tilde{R}_0]_{13}$, 
$[\tilde{R}_1]_{13}$ and $[\tilde{R}_{-1}]_{13}$ using the 
parameters Eq.~(\ref{simplepar})
as well as $\Delta=0$ and obtain
\begin{subequations}
\begin{align}
[\tilde{R}_0]_{13}\,e^{i\Phi} &= \frac{-i \gamma g_{32}^\ast \, 
g_{31} \, g_{42}}
{2 D} e^{-\vec{K}\vec{r} + \phi_0} \,,
\\[2ex]
 g_{41} \, [\tilde{R}_{1}]_{13} &=
 \frac{i \gamma^3 \,  
 ( |g_{31}|^2 + |g_{32}|^2 + |g_{42}|^2)\,|g_{32}|^2\,g_{41}}
{2 D^2} \,,
\\[2ex]
 g_{41}^\ast \, [\tilde{R}_{-1}]_{13}\,e^{2i\Phi} &=
 \frac{- i \gamma \,  
 ( |g_{31}|^2 \, |g_{32}|^2 \, |g_{42}|^2)\,g_{41}^\ast}
{2 D^2}e^{-2\vec{K}\vec{r} + 2\phi_0} \,.
\end{align}
\end{subequations}
\begin{figure}[t]
\includegraphics[width=7cm]{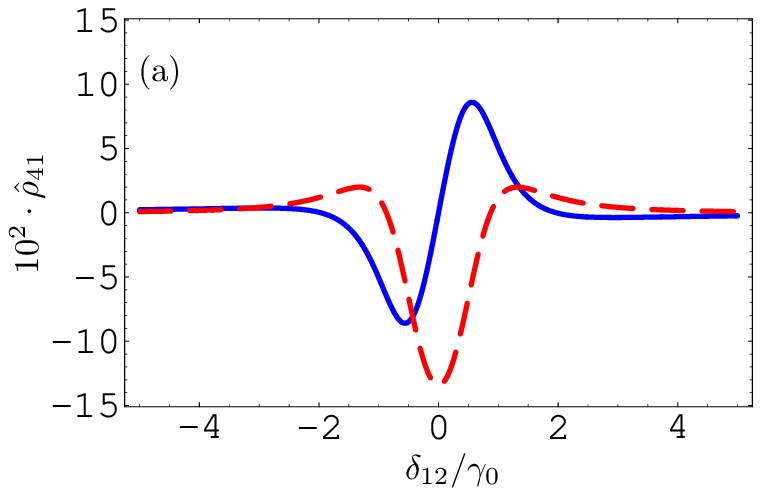}
\includegraphics[width=7cm]{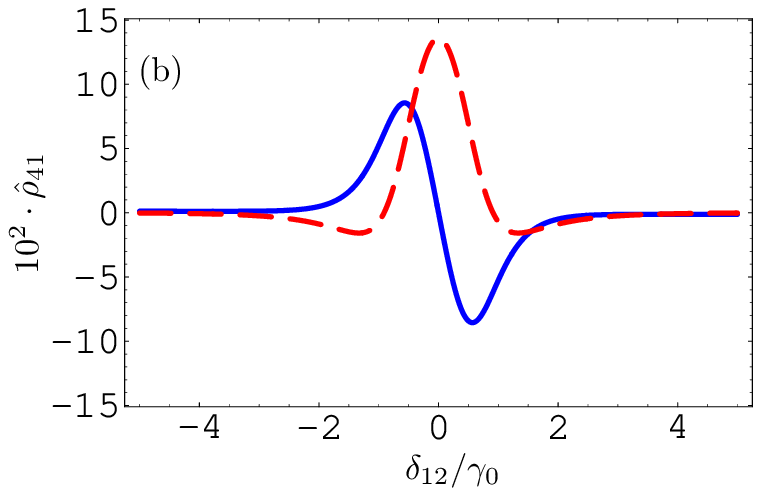}
\includegraphics[width=7cm]{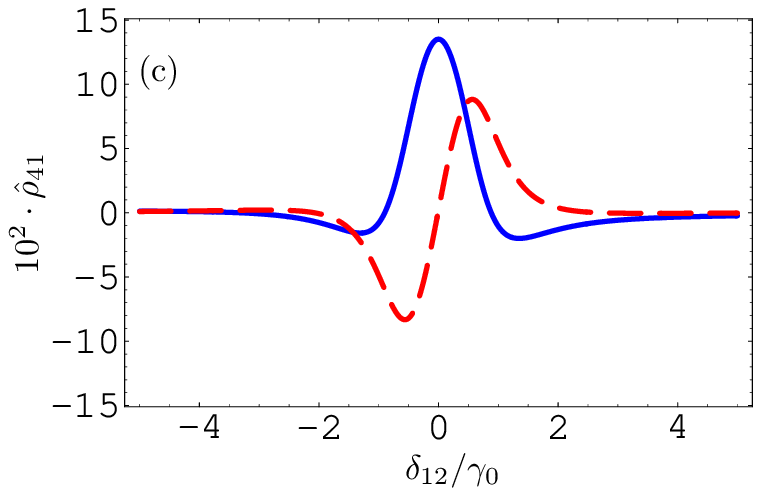}
\caption{\label{fig-resonance}(Color online)
Real (blue solid line) and imaginary (red dashed line)
parts of the atomic density matrix element
$\hat{\rho}_{41}$ in a reference frame oscillating in phase 
with the probe field as a function of the Raman detuning
$\delta_{12}=\Delta_{31}-\Delta_{32}=\Delta_{41}-\Delta_{42}$.
The real part corresponds to the probe field dispersion, whereas
the imaginary part describes the absorptive properties.
The relative initial phase difference between the four applied
laser fields is $\phi_0=0$ in (a), $\phi_0=\pi$ in (b) and
$\phi_0=\pi/2$ in (c).
The common parameters are
$2\gamma_{13}=2\gamma_{14}=2\gamma_{23}=2\gamma_{24}=\gamma_0$, and thus 
$\Gamma_{13}=\Gamma_{14}=\Gamma_{23}=\Gamma_{24}=\gamma_0$, 
$\Gamma_{12}=0.0$, and $\Gamma_{34}=2\gamma_0$.
The detunings are $\Delta_{32}=\Delta_{42}=0$, and the 
Rabi frequencies are chosen as $g_{31}=g_{32}=g_{42}=0.6\gamma_0$ and
$g_{41}=0.01\gamma_0$.}
\end{figure}
Note that $D$ is defined in Eq.~(\ref{def-d}).
Thus the time-dependent Floquet decomposition 
in the limit $\Delta \to 0$, $\vec{K}\to 0$ yields exactly the result
Eq.~(\ref{ana}) of the time-independent analysis, as expected.
The first part of Eq.~(\ref{ana}) representing the scattering
of the driving fields into the probe field mode arises from
$[\tilde{R}_0]_{13}$, as shown in Fig.~\ref{fig-path}(a).
This contribution in general does not oscillate
at the probe field frequency, but rather at the combination 
frequency $\omega_{31}+\omega_{42}-\omega_{32}$ 
of the three driving fields. This frequency
coincides with the probe field frequency only under multiphoton resonance. 
The contribution proportional to $[\tilde{R}_{1}]_{13}$ 
shown in Fig.~\ref{fig-path}(b) is 
in phase with the probe field for all 
values of $\Delta$, and is independent of the relative 
field phase. It represents the direct scattering of the
probe field off of the probe field transition.
The third contribution proportional to $[\tilde{R}_{-1}]_{13}$
can be interpreted as a counter-rotating term which in 
the Floquet expansion differs by $2 \Delta$ from the 
probe field frequency, and is depicted in Fig.~\ref{fig-path}(c).

\begin{figure}[t]
\includegraphics[width=7cm]{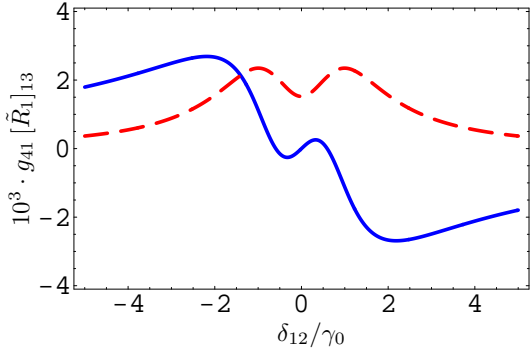}
\caption{\label{fig-rp}(Color online)
Real (blue solid line) and imaginary (red dashed line)
parts of the  contribution to the Floquet decomposition representing
the direct scattering of the probe field off of the probe transition,
$g_{41}\,[\tilde{R}_1]_{13}$ [see Eq.~(\ref{rho-41})].
This figure depicts one of the processes contributing 
to the results in Fig.~\ref{fig-resonance},
and the parameters are the same as in this figure. Note that this particular
contribution is independent of the phase $\phi_0$, such that only one subfigure is shown,
consistent with the interpretation in Sec.~\ref{beyond}.}
\end{figure}
\begin{figure}[t]
\includegraphics[width=7cm]{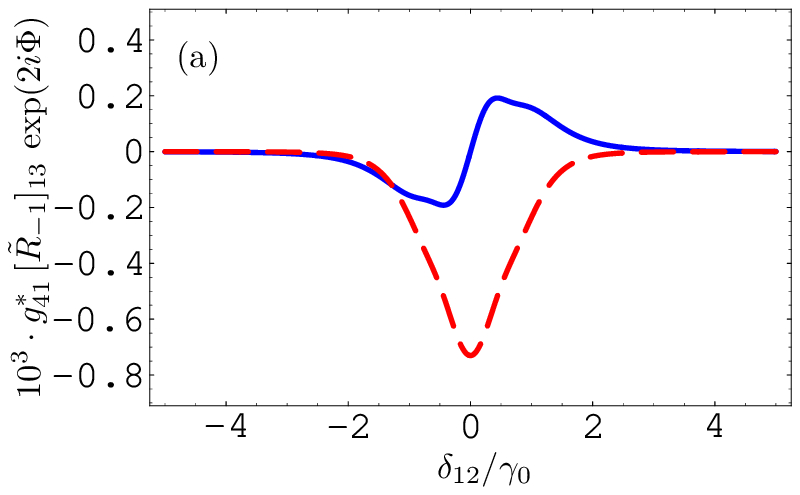}
\includegraphics[width=7cm]{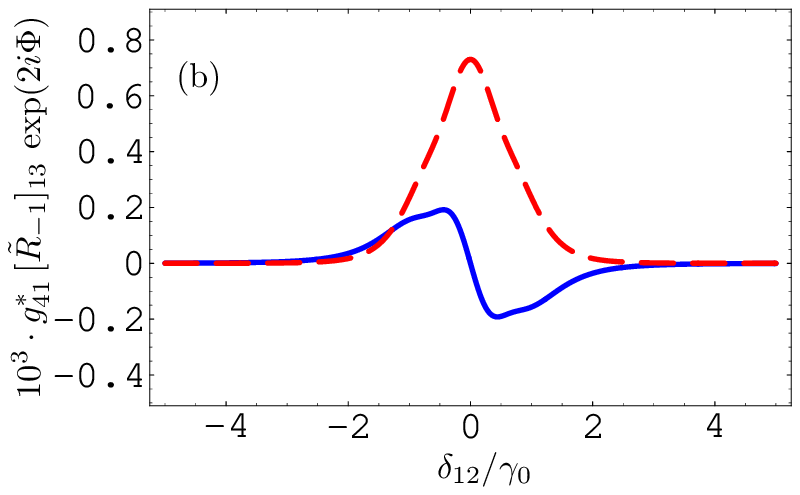}
\caption{\label{fig-rm}(Color online)
Real (blue solid line) and imaginary (red dashed line)
parts of the contribution to the Floquet decomposition representing
a counter-rotating contribution,
$g_{41}^\ast\,[\tilde{R}_{-1}]_{13}\,\exp(2 i \Phi)$ 
[see Eq.~(\ref{rho-41})].
This figure depicts one of the processes contributing to the 
results in Fig.~\ref{fig-resonance},
and the parameters are the same as in this figure. This particular
contribution depends on the phase $2\phi_0$, such subfigure (a) shows 
the result for
$\phi_0 = 0, \pi$ and (b) for $\phi=\pi/2$,
consistent with the interpretation in Sec.~\ref{beyond}.}
\end{figure}

As an important result, we thus conclude that 
the phase-dependence of the loop-configuration
studied here is restricted to the multiphoton resonance
condition $\Delta=0$, because it arises from the scattering
of the coupling fields into the probe field mode. 
Furthermore, it can be seen that all contributions
but the direct scattering acquire an additional dependence on 
the wave vector mismatch $\vec{K}$, which influences the spatial
emission pattern of these contributions. In general, only the direct 
scattering contribution can be detected in propagation direction
of the probe beam regardless of the separation of detector and the
scattering atoms.
Finally, we note that the first expression
$[\tilde{R}_0]_{13}$ is independent of $\Delta$, whereas 
$[\tilde{R}_{1}]_{13}$ and $[\tilde{R}_{-1}]_{13}$
have complicated dependencies on $\Delta$.


\section{\label{sec:res}Results and discussion}
We now turn to the numerical study of our results from
the master equations~(\ref{density}).
In a first step we assume that the multiphoton resonance condition 
$\Delta = 0, \vec{K}=0$ is fulfilled. Then the steady state 
solution for Eqs.~(\ref{density}) exist
and depends on the initial constant relative phase $\phi_0$ of the 
driving fields [see, e.g., Eq.~(\ref{ana}) for the special case of parameters
Eq.~(\ref{simplepar})]. In Fig.~\ref{fig-resonance} we plot 
the absorption and dispersion 
spectrum versus the Raman detuning 
$\delta_{12}=\Delta_{31}-\Delta_{32}=\Delta_{41}-\Delta_{42}$, 
for different initial phases $\phi_0$. The
common parameters are chosen as 
$2\gamma_{13}=2\gamma_{14}=2\gamma_{23}=2\gamma_{24}=\gamma_0$, 
where $\gamma_0$ is the total decay rate of one of the upper
transitions to a lower transition.
Then, $\Gamma_{13}=\Gamma_{14} =\Gamma_{23}=\Gamma_{24}=\gamma_0$, 
$\Gamma_{12}=0$, and $\Gamma_{34}=2\gamma_0$.
Further, $\Delta_{32}=\Delta_{42}=0$, and the driving fields
have Rabi frequencies $g_{13}=g_{23}=g_{24}=0.6\gamma_0$, 
$g_{14}=0.01\gamma_0$.
The initial phase values in Fig.~\ref{fig-resonance} are $\phi_0=0$ in (a), 
$\phi_0=\pi$ in (b) and $\phi_0=\pi/2$ in (c). 
The slope of the dispersion (real part) is positive for
$\phi_0=0$, and accompanied by gain (negative imaginary part). 
For $\phi_0=\pi$, however, the slope of the dispersion becomes
negative together with an absorption peak. 
For $\phi_0=\pi/2$, the curve shapes of real and imaginary part
are interchanged. The results
show the phase-dependence of the probe absorption for a 
closed-loop configuration
in good agreement with previous results as in Ref.~\cite{Korsunsky}.

We now evaluate the individual contributions to the Floquet 
decomposition of the density matrix, Eqs.~(\ref{rho-41}),
using the same parameters as above in Fig.~(\ref{fig-resonance}).
From Eq.~(\ref{rho-41}), it is obvious that the numerically
dominant contribution arises from $[\tilde{R}_0]_{13}\,\exp(i\Phi)$,
since the two other contributions are suppressed by one power
of the weak driving field Rabi frequency $g_{41}$. Consistent
with this interpretation, the results for $[\tilde{R}_0]_{13}\,\exp(i\Phi)$
are almost identical to the ones shown in Fig.~\ref{fig-resonance},
with the difference given by the two other terms in Eq.~(\ref{rho-41})
and the neglected higher-order terms of order $\mathcal{O}(g_{41}^2)$.
Therefore, in the following, we only show the two difference contributions 
$g_{41} [\tilde{R}_1]_{13}$ and $g_{41}^\ast [\tilde{R}_{-1}]_{13}\,
\exp(2 i\Phi)$.
Fig.~\ref{fig-rp} shows the contribution $g_{41} [\tilde{R}_1]_{13}$, 
which is the process contributing 
to the results in Fig.~\ref{fig-resonance} depicted in Fig.~\ref{fig-path}(b).
As discussed in Sec.~\ref{beyond}, this process is insensitive to the
phase $\phi_0$, such that only one subfigure is shown for all values of 
$\phi_0$. Note that 
the results confirm our expectation that the numerical contribution 
of this process to the full results in Fig.~\ref{fig-resonance} is small.
The contribution $g_{41}^\ast [\tilde{R}_{-1}]_{13}\,\exp(2i\Phi)$ 
is shown in Fig.~\ref{fig-rm}; the corresponding process is depicted in 
Fig.~\ref{fig-path}(c).
This process depends on  $2 \phi_0$, such that subfigure (a) shows the
identical results for $\phi_0 = 0$ and $\phi_0 = \pi$, whereas
subfigure (b) shows the result for $\phi_0 = \pi/2$. Also the numerical 
contribution 
of this process to Fig.~\ref{fig-resonance} is small as compared to 
$[\tilde{R}_0]_{13}\,\exp(i\Phi)$.
As discussed in Secs.~\ref{intro} and \ref{suscep}, 
the slope of the dispersion 
spectrum against the probe field detuning has a major role in 
the calculation of the group velocity. But to define 
the group velocity of a probe pulse in a given medium properly,
one has to require that the medium itself including the coupling 
fields has properties independent of the probe pulse 
characteristics. Note that in this context the coupling fields
prepare the medium, and in that sense belong to the medium
in which the probe pulse propagates, and thus their detunings
needs to be kept fixed.
The probe pulse, however, has a finite duration and thus consists 
of different frequency components  which 
interact with the atom with different detunings simultaneously.
In particular, it is not possible to, e.g., adjust one 
of the coupling field
detunings such that the multiphoton resonance condition is maintained
throughout the pulse propagation. Therefore, the propagation of a pulse
cannot be described assuming the multiphoton resonance condition,
and our numerical results Fig.~\ref{fig-resonance}-\ref{fig-rm} 
presented so far are not sufficient to evaluate the group velocity.

\begin{figure}[t]
\includegraphics[width=7cm]{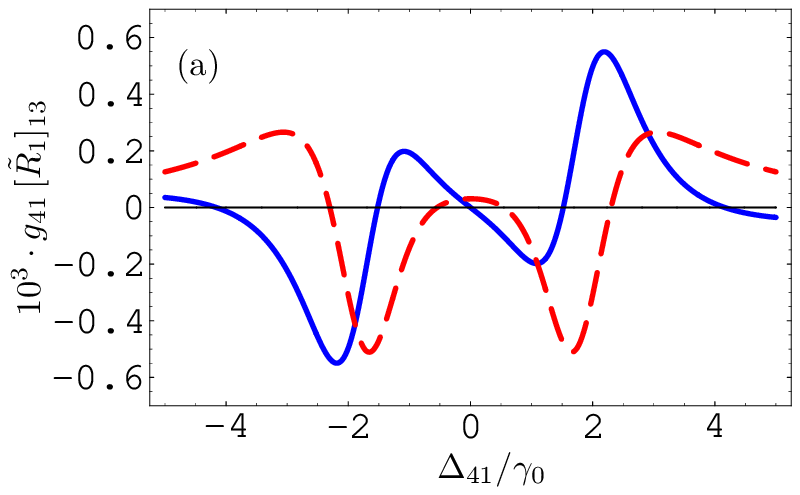}
\includegraphics[width=7cm]{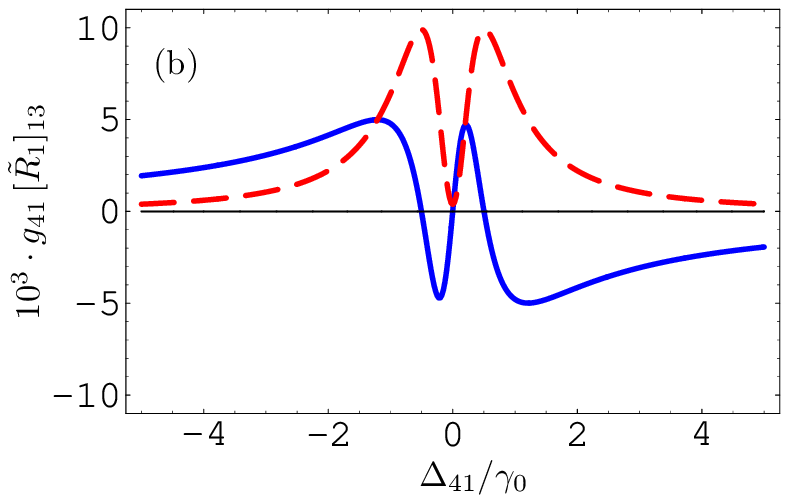}
\caption{\label{fig-loop-timedep}(Color online)
Real (blue solid line) and imaginary (red dashed line)
parts of the contribution to the Floquet decomposition representing
the direct scattering of the probe field off of the probe transition,
$g_{41}\,[\tilde{R}_1]_{13}$ [see Eq.~(\ref{rho-41})].
The susceptibility is plotted against the probe field
detuning $\Delta_{41}$. This corresponds to a calculation 
relevant for the 
evaluation of the group velocity, but violates the
multiphoton resonance condition and thus requires
a Floquet analysis.
The parameters in (a) are
$\Delta_{31}=\Delta_{32}=\Delta_{42}=0$,
$2\gamma_{13}=2\gamma_{14}=2\gamma_{23}=2\gamma_{24}=\gamma_0$, 
$g_{31}=1.8 \gamma_0$, $g_{32} = 0.2 \gamma_0$, $g_{42}=0.5\gamma_0$, and
$g_{41}=0.01\gamma_0$.
Subfigure (b) shows the case
$\Delta_{31}=10\gamma_0$, $\Delta_{32}=\Delta_{42}=0$,
$2\gamma_{13}=2\gamma_{14}=2\gamma_{23}=2\gamma_{24}=\gamma_0$, 
$g_{31}=g_{32} = 0.1 \gamma_0$, $g_{42}=0.5\gamma_0$, and
$g_{41}=0.01\gamma_0$.
}
\end{figure}

\begin{figure}[t]
\includegraphics[width=7cm]{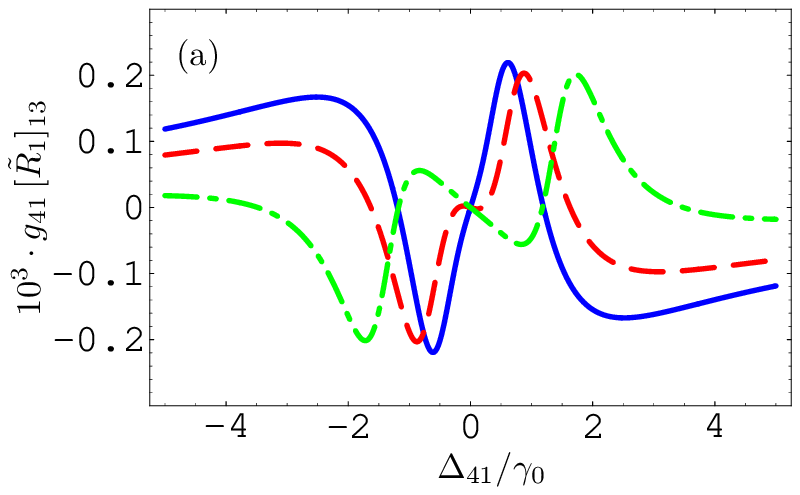}
\includegraphics[width=7cm]{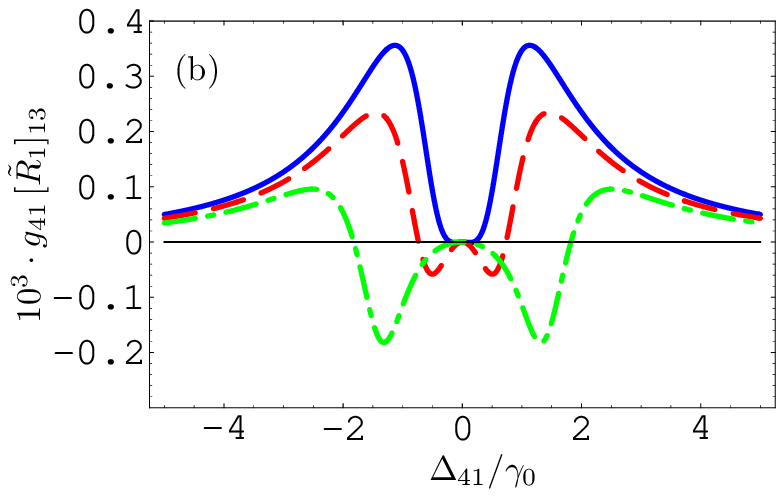}
\caption{\label{fig-no-loop}(Color online)
Real (a) and imaginary (b)
parts of the scaled contribution to the Floquet decomposition representing
the direct scattering of the probe field off of the probe transition,
$[\tilde{R}_1]_{13}$ [see Eq.~(\ref{rho-41})].
The susceptibility is plotted against the probe field
detuning $\Delta_{41}$ with other detunings chosen as 
$\Delta_{31}=\Delta_{42}=0$. 
One of the Rabi frequencies is chosen as zero, $g_{32}=0$,
such that the setup corresponds to the one shown in 
Fig~\ref{fig-system}(b).
The solid blue line is for $g_{31}= 0.7 \gamma_0$, 
the red dashed line for $g_{31}= 0.85 \gamma_0$,
and the green dash-dotted line for $g_{31}= 1.5 \gamma_0$.
The other parameters are 
$2\gamma_{13}=2\gamma_{14}=2\gamma_{23}=2\gamma_{24}=\gamma_0$, 
$g_{42}= 0.2 \gamma_0$, and $g_{41}=0.01\gamma_0$.}
\end{figure}
Therefore, in Fig.~\ref{fig-loop-timedep}(a), we show the
probe field susceptibility against the probe field detuning
$\Delta_{41}$, while the coupling field detunings are 
$\Delta_{31}=\Delta_{32}=\Delta_{42}=0$.
This corresponds to a calculation relevant for the 
evaluation of the group velocity, but violates the
multiphoton resonance condition $\Delta = 0$ for most values
of $\Delta_{41}$ shown in the figure and thus 
requires the use of the time-dependent Floquet analysis.
Consequently, in Fig.~\ref{fig-loop-timedep}, we only show the
component $[\tilde{R}_1]_{13}$ since this part of the
Floquet decomposition is the only one which oscillates
in phase with the probe field under these conditions.
It can be seen in Fig.~\ref{fig-loop-timedep}(a) that 
around $\Delta_{41}=\pm 2 \,\gamma_0$, the real part
of the susceptibility has positive slope, while the imaginary
part is strongly negative. This indicates
subluminal light propagation with gain.
At about $\Delta_{41}=0$, the real part
of the susceptibility has negative slope over a wide frequency
range, together with a small positive or even negative imaginary
part. This corresponds to superluminal propagation with small
absorption or gain.

If, however, the parameters and the laser field alignments
are such that for a particular value of
$\Delta_{41}$ within the resonance structure in the susceptibility
the multiphoton resonance condition is fulfilled, then 
the coupling fields may be scattered into the probe field mode.
This would give rise to a huge, very narrow contribution to 
the probe field susceptibility [see Fig.~\ref{fig-resonance}], 
which is likely to completely distort the probe pulse shape.
In Fig.~\ref{fig-loop-timedep}(a), this could occur at $\Delta_{41}=0$
if the laser field wavevectors are chosen suitably.
Therefore for probe pulse propagation, parameters
may be favorable which suppress the scattering of the coupling fields
into the probe field mode. But under such conditions, 
the system is not phase-dependent.

In Fig.~\ref{fig-loop-timedep}(b), we show a case where 
the multiphoton resonance condition $\Delta = 0$ is 
violated for all shown values of $\Delta_{41}$.
This avoids the scattering of the coupling fields
into the probe field mode, and 
is accomplished by choosing the detunings as
$\Delta_{31}=10 \,\gamma_0$, and $\Delta_{32}=\Delta_{42}=0$.
Around $\Delta_{41}=0$, subluminal light propagation 
with little absorption can be achieved.
The fact that this and similar result can be obtained
for a rather large detuning $\Delta_{31}=100\,g_{31}$ 
as compared to the corresponding Rabi frequency
then suggest that a closed-loop
configuration of the laser fields may not be necessary 
at all.

Therefore, in the final part, we discuss  the dispersion and
absorption spectrum for the system without a closed interaction 
loop shown in Fig.~\ref{fig-system}(b).
Fig.~\ref{fig-no-loop} shows again the probe field susceptibility
due to the direct scattering of the probe beam off of the
probe transitions, with Rabi frequency $g_{32}=0$. Subfigure
(a) shows the real part, (b) the imaginary part.
The solid blue line is for $g_{31}= 0.7 \gamma_0$, 
the red dashed line for $g_{31}= 0.85 \gamma_0$,
and the green dash-dotted for $g_{31}= 1.5 \gamma_0$.
All coupling field detunings are zero, 
and the other parameters are 
$2\gamma_{13}=2\gamma_{14}=2\gamma_{23}=2\gamma_{24}=\gamma_0$, 
$g_{42}= 0.2 \gamma_0$, and $g_{41}=0.01\gamma_0$.
If the Rabi frequency $g_{31}$ for transition $|1\rangle-|3\rangle$ 
is small, then the results are similar to the simple
three-level electromagnetically induced transparency (EIT) case. 
The slope of the dispersion is positive, and subluminal light
propagation appears with an EIT dip in the absorption spectrum,
see the solid blue line in Fig.~\ref{fig-no-loop}. 
With increasing Rabi frequency, the slope of
the dispersion around zero detuning becomes negative and 
superluminal light propagation sets in. Together with
superluminal light propagation, zero absorption
or even gain is achieved. 
Thus the intensity of the pump fields can be used as
a simple control parameter to switch the light propagation 
from subluminal to superluminal.

\begin{figure}[t!]
\includegraphics[width=7cm]{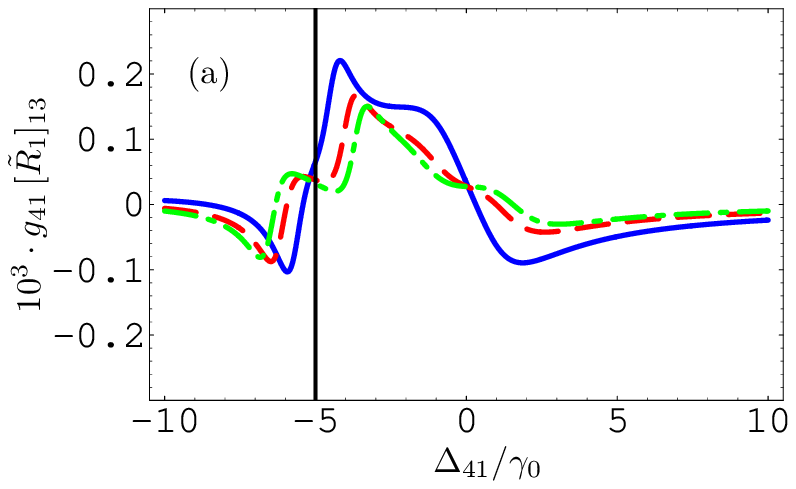}
\includegraphics[width=7cm]{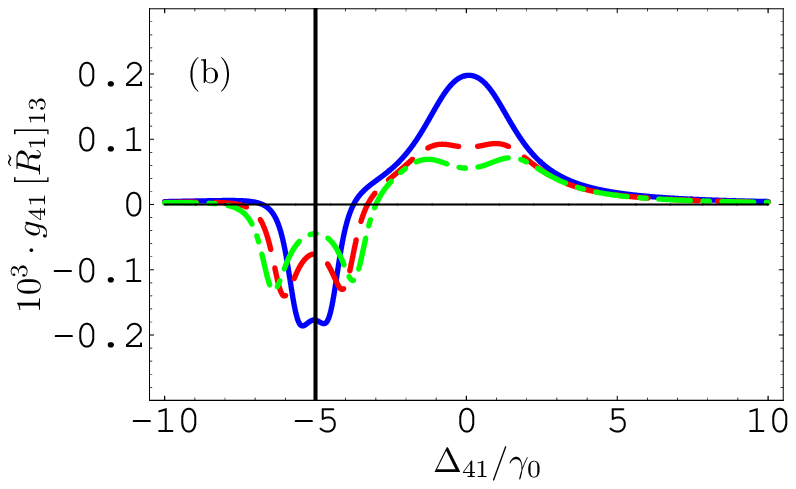}
\caption{\label{fig-no-loop2}(Color online)
As in Fig.~\ref{fig-no-loop}, but with parameters
$g_{31}= 0.7 \gamma_0$ for the solid blue line, 
$g_{31}= 0.85 \gamma_0$ for the  red dashed line,
and $g_{31}= 0.85 \gamma_0$ for the green dash-dotted line.
The other parameters are 
$\Delta_{42}=-5\gamma_0$, $\Delta_{31}=0$,  $g_{32}=0$,
$2\gamma_{13}=2\gamma_{14}=2\gamma_{23}=2\gamma_{24}=\gamma_0$, 
$g_{42}= 0.6 \gamma_0$, and $g_{41}=0.01\gamma_0$.
The solid vertical line indicates $\Delta_{41}=4\gamma_0$.}
\end{figure}

Fig.~\ref{fig-no-loop2} shows another set of results
for the system in Fig.~\ref{fig-system}(b), with parameters
$\Delta_{42}=-5\gamma_0$, $\Delta_{31}=0$,  $g_{32}=0$,
$2\gamma_{13}=2\gamma_{14}=2\gamma_{23}=2\gamma_{24}=\gamma_0$, 
$g_{42}= 0.6 \gamma_0$, and $g_{41}=0.01\gamma_0$,
and $g_{31}= 0.7 \gamma_0$ for the solid blue line, 
$g_{31}= 0.85 \gamma_0$ for the  red dashed line,
and $g_{31}= 0.85 \gamma_0$ for the green dash-dotted line.
Due to the different coupling field detuning $\Delta_{42}$,
the interesting probe susceptibility structure is shifted
to $\Delta_{41}\approx -5\gamma_0$. In this region,  
for $g_{13}=0.8\gamma$, one finds subluminal light propagation 
accompanied by strong gain. Increasing the Rabi frequency
$g_{13}$, again the light propagation switches to superluminal,
still together with gain.

\section{\label{sec:conc}Discussion and summary}

We have discussed the propagation of a probe pulse
through a medium that is driven by coupling and probe
laser fields that form a closed interaction loop.
This gives rise to a dependence of the system on the
relative phase of the various laser fields, but in
general prohibits the existence of a time-independent
steady state of the system. A stationary steady state
only exists if the so-called multiphoton
resonance condition $\Delta = 0$ [see Eq.~(\ref{phi})]
is fulfilled. A probe pulse, however, has a finite
frequency width. Therefore, a time-independent analysis
is insufficient to correctly describe the pulse
propagation through the medium. To solve this problem,
we have solved the time-dependent system without assuming
the multiphoton resonance condition by means of
a Floquet decomposition of the equations of motion.
We found that the different Floquet components can be 
interpreted in terms of different scattering
processes. The phase dependence arises from a scattering
of the coupling fields into the probe field mode, and
thus only occurs at a specific probe field frequency.
If this frequency is within the frequency width of 
the probe pulse, a strong distortion of the pulse
shape can be expected. In some cases, it may be possible
to find probe pulses that are sufficiently narrow
in the frequency domain such that they fit into the
frequency range where the scattering of the coupling fields
into the probe field mode dominates. Especially for shorter
and thus broader pulses, however, we conclude that it may
be advantageous to avoid the scattering of the coupling
fields into the probe field mode. Then, however, the
system is no longer phase dependent, and a closed 
interaction loop may not be necessary at all.
We expect that similar results hold for other
level schemes which involve a closed laser field interaction
loop.

Apart from these general considerations, using our time-dependent
Floquet analysis, we have shown that for realistic parameter
sets both the closed-loop 
double-$\Lambda$ system and the corresponding system without 
a closed loop allow for sub- and superluminal light propagation
with small absorption or even with gain.
Further, we have identified the Rabi frequency of one of the
coupling fields as a convenient control parameter
to switch the light propagation between 
sub- and superluminal light propagation.

\begin{acknowledgments}
MM gratefully acknowledges support for this work from the German
Science Foundation and from
Zanjan University. He also thanks M. Sahrai for useful
discussions.
\end{acknowledgments}


\end{document}